\documentclass[a4paper,fleqn,usenatbib]{mnras}

\usepackage[T1]{fontenc}
\usepackage{ae,aecompl}
\usepackage{ulem} 

\usepackage{graphicx}	
\usepackage{amsmath}	
\usepackage{amssymb}	
\usepackage{float}
\usepackage{dblfloatfix}
\usepackage{subfigure}
\newcommand{\eagle}{{\sc eagle}}

\title[\eagle\ Mass-Metallicity-Kinematics Relation]
{Correlations between mass, stellar kinematics and gas metallicity in \eagle\ galaxies}

\author[Zenocratti et al.]{
L. J. Zenocratti,$^{1}$\thanks{E-mail: ljzenocratti@hotmail.com}
M. E. De Rossi,$^{2,3}$\thanks{E-mail: mariaemilia.dr@gmail.com}
M. A. Lara-L\'opez,$^{4}$
T. Theuns$^{5}$
\\
$^{1}$Facultad de Ciencias Astron\'omicas y Geof\'isicas, Universidad Nacional de La Plata, Paseo del Bosque s/n, B1900FWA, La Plata, Argentina\\
$^{2}$Universidad de Buenos Aires, Facultad de Ciencias Exactas y Naturales y Ciclo B\'asico Com\'un. Buenos Aires, Argentina\\
$^{3}$CONICET-Universidad de Buenos Aires, Instituto de Astronom\'ia y F\'isica del Espacio (IAFE). Buenos Aires, Argentina\\ 
$^{4}$DARK, Niels Bohr Institute, University of Copenhagen, Lyngbyvej 2, Copenhagen DK-2100, Denmark\\
$^{5}$Institute for Computational Cosmology, Physics Department, University of Durham, South Road, Durham DH1 3LE, UK \\
}

\date{Accepted XXX. Received YYY; in original form ZZZ}

\pubyear{2019}

\begin{document}
\label{firstpage}
\pagerange{\pageref{firstpage}--\pageref{lastpage}}
\maketitle

\begin{abstract}
The metallicity of star-forming gas in galaxies from the \eagle\ simulations increases with stellar mass. Here we investigate whether the scatter around this relation correlates with morphology and/or stellar kinematics. At redshift $z=0$, galaxies with more rotational support have lower metallicities on average when the stellar mass is below $M_\star\approx 10^{10}~{\rm M}_\odot$. This trend inverts at higher values of $M_\star$, when prolate galaxies show typically lower metallicity. At increasing redshifts, the trend between rotational support and metallicity becomes weaker at low stellar mass but more pronounced at high stellar mass. We argue that the secondary dependence of metallicity on stellar kinematics is another manifestation of the observed anti-correlation between metallicity and star formation rate at a given stellar mass. At low masses, such trends seem to be driven by the different star-formation histories of galaxies and stellar feedback. At high masses, feedback from active galactic nuclei and galaxy mergers play a dominant role.
\end{abstract}




\section{Introduction}
\label{sec:intro}

The relation between stellar mass and gas-phase metallicity in galaxies (henceforth the mass-metallicity relation, MZR) has been studied extensively in the last decades from both an observational 
(\citealp{Tremonti2004}; \citealp{laralopez2010}) and a theoretical (\citealp{Calura2009}; \citealp{Yates2012}; \citealp{derossi2015}; \citealp{DeRossi2017}-hereafter, DR17; \citealp{sharma2019}) point of view. At redshift $z\sim 0$, gas metallicity, $Z$, increases with stellar mass, $M_\star$, approximately as a power-law, $Z\propto M_\star^{2/5}$; the slope of the correlation flattens towards higher masses. This power-law trend is also seen at higher redshifts, though possibly with a different slope and normalisation \cite[e.g.][]{troncoso2014}.

The scatter along the observed MZR correlates with other properties of galaxies. \citet{Ellison2008} showed that, at given stellar mass, observed galaxies with smaller half-mass radii 
or lower specific star formation rates (sSFR) tend to have higher gas metallicity,
as quantified by the oxygen abundance, O/H. To account for these observations, \citet{laralopez2010} and  \citet{Mannucci2010} suggested the existence of a three-dimensional relation between $M_\star$, O/H and star formation rate (SFR), where systems with higher SFRs tend to have lower O/H at a given value of $M_\star$. Alternatively, this 3D relation could result from a more fundamental underlying relation between $M_\star$, O/H and {\it gas fraction} ($f_{\rm g}$), since $f_{\rm g}$ and SFR correlate (e.g. \citealt{Bothwell2013}, \citealp{laralopez2013c}). Deciding which relation is more fundamental could be helped by examining other correlations. Recent observations suggest that $Z$ tends to be lower in galaxies with a higher concentration, higher S\'ersic index, or higher SFR \citep{Wu2019}, at given value of $M_\star$. Unfortunately, surprisingly large uncertainties remain in inferring physical relations from the data because different methods yield significantly different answers \citep[e.g.][]{telford2016}. In fact, some observational studies do not find that the scatter around the MZR correlates with SFR \citep[e.g.][]{sanchez2019}, and some studies claim that the correlation exists but inverts at high $M_\star$ \citep[e.g.][]{Yates2012}.

In this paper we examine the scatter around the MZR in galaxies from the \eagle\ cosmological hydrodynamical simulations \citep{Schaye2015}. The \lq Evolution and Assembly of GaLaxies and their Environments\rq\ (\eagle) suite of cosmological hydrodynamical simulations uses sub-grid models calibrated to reproduce a small set of observations at $z\approx0$, as described in \cite{Crain2015}. The simulations then reproduce a relatively extensive set of other observations, including the evolution of the galaxy stellar mass function \citep{Furlong2015}, of galaxy sizes \citep{Furlong2017}, of optical \citep{Trayford2015} and UV and IR luminosities \citep{Camps2018}. DR17 analysed the secondary metallicity dependencies in {\eagle} (including the dependencies on SFR, sSFR, $f_{\rm g}$ and stellar age, ${t}_{\star}$), obtaining good agreement with observed trends \citep[see also][]{laralopez2019}. In particular, DR17 show that \eagle\ galaxies follow remarkably well the observed \lq Fundamental Metallicity Relation\rq\ introduced by \cite{Mannucci2010}. In addition, \citet{Sanchez2018} show that \eagle\ simulations are able to reproduce the observed secondary metallicity dependence on the size of galaxies.

In this Letter, we report new predictions of \eagle\ simulations regarding the connection 
between MZR scatter and internal morpho-kinematics of galaxies. 
Such trends were not previously reported in MZR studies.
In a forthcoming article, we address the origin of these metallicity secondary dependencies
by analysing the formation histories of different galaxy populations.
This Letter is organized as follows. In Section \ref{sec:simulation}, we briefly describe the \eagle\
simulations and the galaxy selection criteria. In Section \ref{sec:mzr_krot}, we analyse
the simulated MZR as a function of the morphology and kinematics of the galaxies. We discuss the origin of our obtained trends in Section \ref{sec:discussion}.
We summarise our findings in Section~\ref{conclusions}. 
      
\section{The EAGLE Simulations}
\label{sec:simulation}
A full description of the \eagle\ simulation suite is given by \cite{Schaye2015}. Briefly, the suite was simulated with the {\sc gadget-3} incarnation of the {\sc treepm-sph} code described by \cite{Springel2005}, with sub-grid modules for physics whose parameters are calibrated to reproduce the $z\approx0$ galaxy stellar mass function, the relation between galaxy mass and size, and the black hole mass - stellar mass relation \citep{Crain2015}. The adopted cosmological parameters are taken from \cite{Planck2015}: $\Omega_{\Lambda}=0.693$, $\Omega_{\rm{m}}=0.307$, $\Omega_{\rm{b}}=0.04825$, $n_{\rm{s}}=0.9611$, $Y=0.248$, and $h=0.677$.

\eagle\ consists of simulations with various box sizes and particle masses. Here, we mainly use simulation labelled \lq Ref-L100N1504\rq\ in \cite{Schaye2015}, which has a co-moving extent of $L=100$~co-moving megaparsecs (cMpc) and a baryonic particle mass of $\sim 1.2\times 10^6~{\rm M}_\odot$ (corresponding to $1504^3$ particles). We have verified that the main trends and conclusions presented in this work are consistent with those from the higher-resolution \eagle\ simulation \lq Recal-L025N0752\rq, analysed previously by DR17.

\eagle\ galaxies are identified using a combination of the \lq friends-of-friends\rq\ ({\sc fof}) and {\sc subfind} 
algorithms. This picks-out \lq self-bound\rq\ structures of gas, stars and dark matter. Here we analyse the properties of both central galaxies (the dominant galaxies in {\sc fof} halos) and satellites\footnote{Our main results are unchanged if we analyse only centrals.}. Following DR17, we measure baryonic properties within spherical apertures of 30 proper kilo-parsecs (pkpc) and characterize the \lq metallicity\rq\ of star-forming gas by its O/H abundance (\eagle\ tracks 11 abundances, including oxygen and hydrogen). We analyse galaxies with at least 25 star-forming gas particles (gas mass at least $5.25\times 10^7~{\rm M}_\odot$) which we found to be a reasonable compromise between numerical resolution and bias. To characterize the stellar morphology and kinematics,
we use the fraction of kinetic energy in co-rotation, $\kappa_{\rm co}$, the disc-to-total stellar mass ratio, $D/T$, the ratio $V/\sigma$ of stellar rotation to velocity dispersion, the ellipticity, $\epsilon_\star$, of the stellar body, and its triaxiality, $T$. These were computed by \citet{Thob2019} and can be queried in the \eagle\ database\footnote{http://eagle.strw.leidenuniv.nl, http://www.eaglesim.org/} (\citealp{mcalpine2016}; \citealp{eagle2017}).

\begin{figure}
\centering
\subfigure{\includegraphics[width=0.38\textwidth]{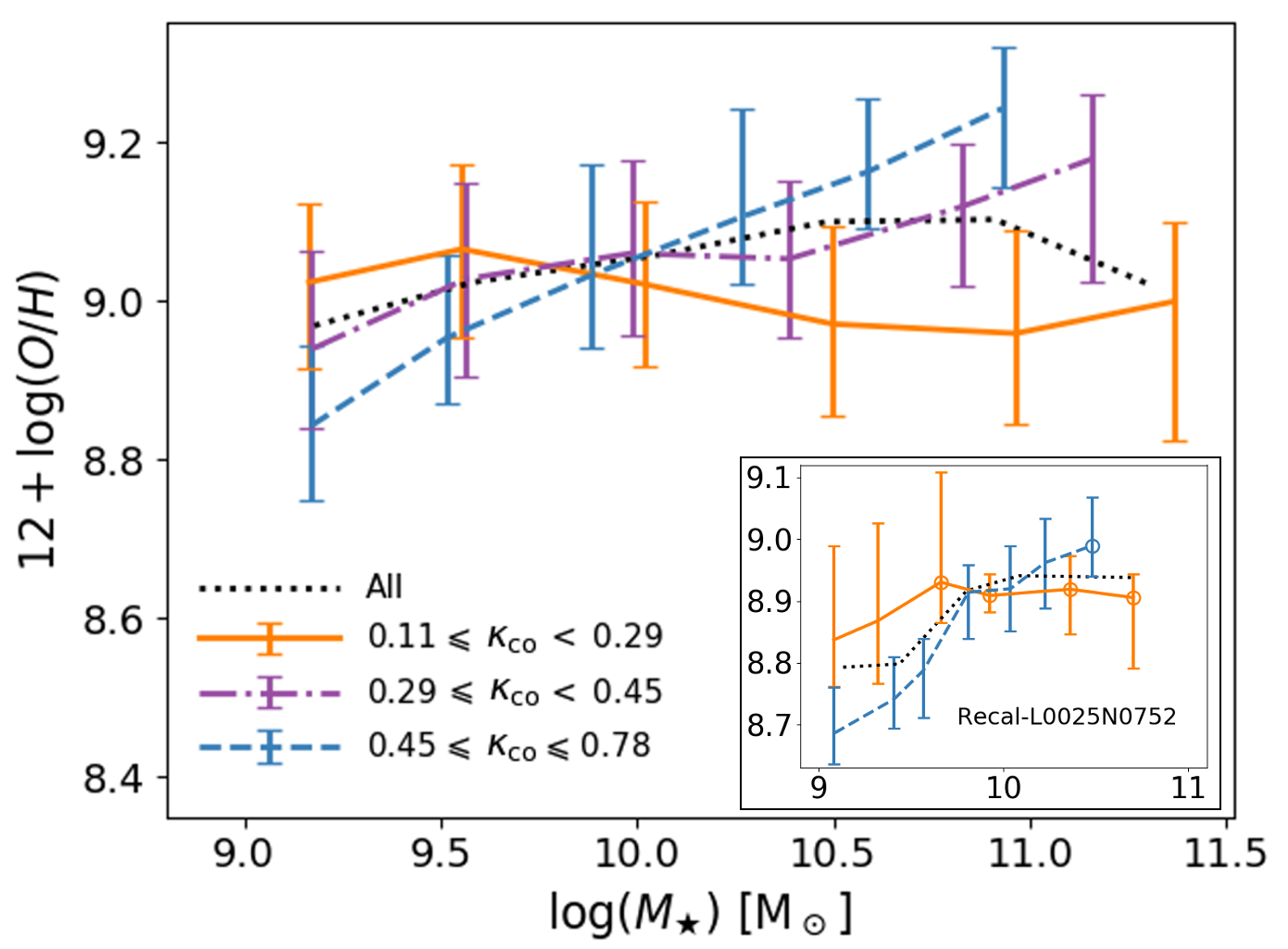}}
	\caption{{\it Black dotted line:} median $M_{\star}$ - O/H (MZR) relation in redshift $z=0$ \eagle\ galaxies from simulation Ref-L100N1504. 	{\it Coloured lines:} MZR relation for \eagle\ galaxies binned by $\kappa_{\rm co}$, the fraction of stellar kinetic energy in rotation: the lowest third $\kappa_{\rm co}$ ({\it orange}), intermediate $\kappa_{\rm co}$ ({\it purple}) and
	the highest third $\kappa_{\rm co}$ ({\it blue}). Error bars encompass the 25$^{\rm th}$ and 75$^{th}$ percentiles. {\it Inset:} as main panel, but for the higher resolution \eagle\ simulation Recal-L0025N0752.}
\label{fig.1}
\end{figure}

\begin{figure*}
\centering
\subfigure{\includegraphics[width=0.32\textwidth]{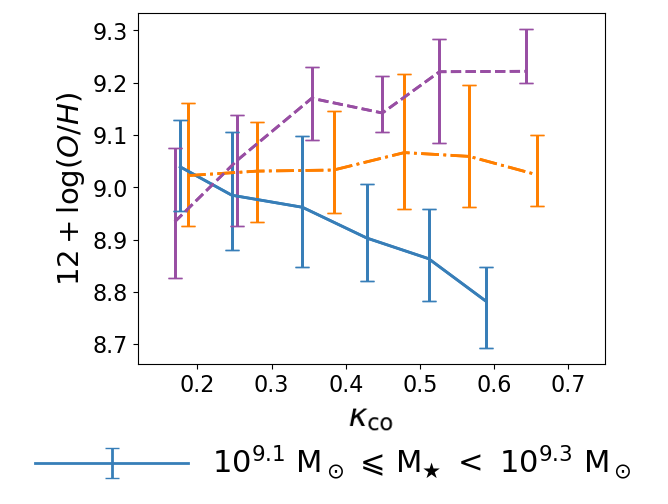}}
	\subfigure{\includegraphics[width=0.32\textwidth]{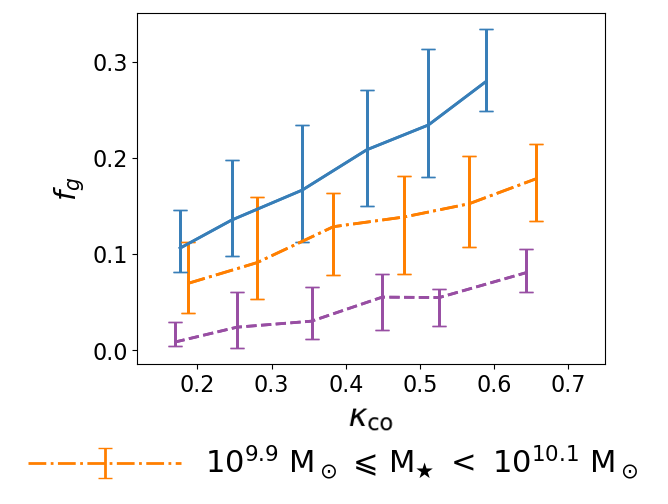}}
\subfigure{\includegraphics[width=0.32\textwidth]{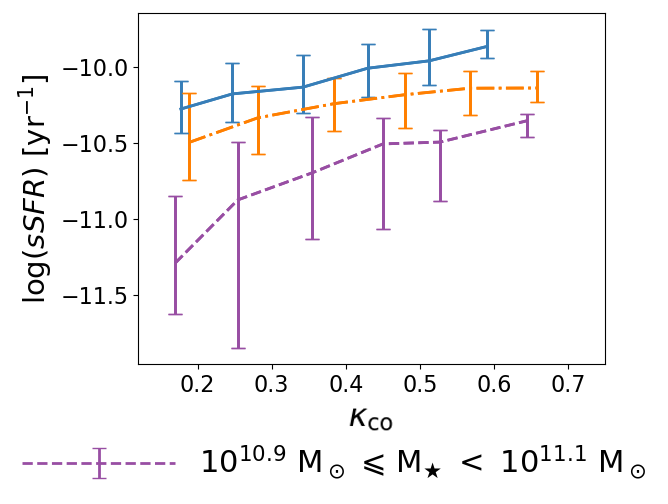}}
	\caption{Correlation between O/H ({\it left panel}), gas fraction $f_g$ ({\it middle panel}), and the specific star formation rate ({\it right panel)} and $\kappa_{\rm co}$, the fraction of stellar kinetic energy in rotation, for $z=0$ \eagle\ galaxies from simulation Ref-L100N1504.
	Galaxies are binned in stellar mass: low stellar mass ({\it blue}), intermediate stellar mass ({\it orange}) and high stellar mass ({\it purple}), see legend. Curves represent the median relation, error bars encompass the 25$^{\rm th}$ and 75$^{\rm th}$ percentiles.}
\label{fig.2}
\end{figure*}

\begin{figure*}
\centering
\subfigure{\includegraphics[width=0.32\textwidth]{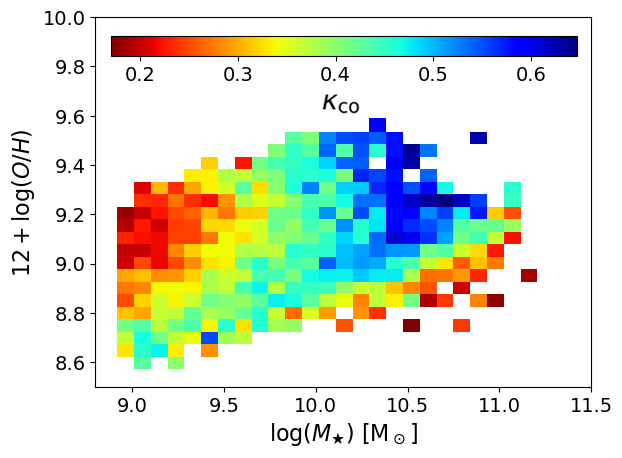}}
\subfigure{\includegraphics[width=0.32\textwidth]{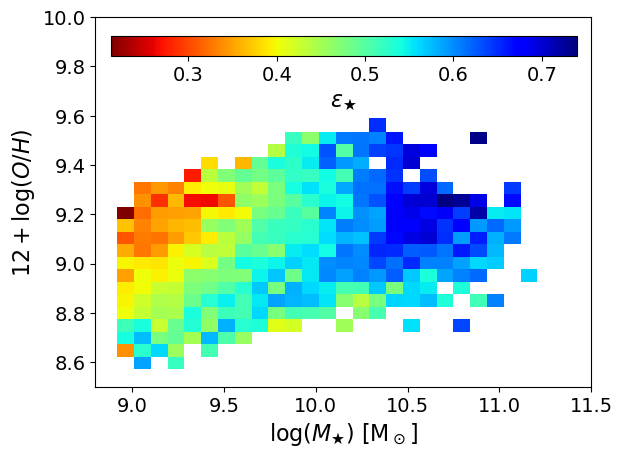}}
\subfigure{\includegraphics[width=0.32\textwidth]{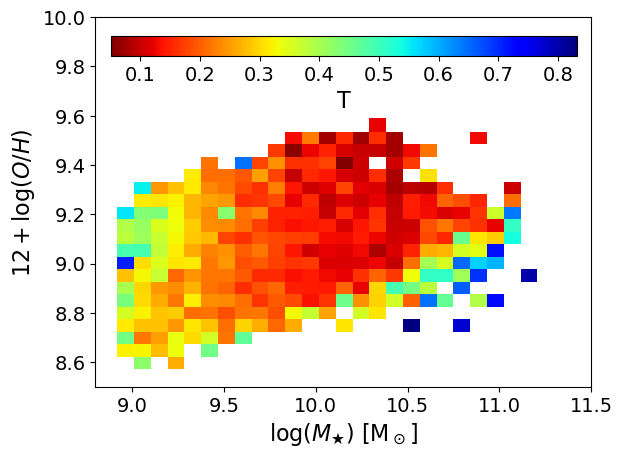}}
	\caption{O/H metallicity as a function of stellar mass, $M_{\star}$, for $z=0$ \eagle\ galaxies from simulation Ref-L100N1504.
	Bins in O/H-$M_\star$ are coloured according to the median value of $\kappa_{\rm co}$ ({\it left panel}), the stellar ellipticity
	$\epsilon_\star$ ({\it middle panel}) and the galaxy's triaxiality parameter $T$ ({\it right panel}).}
\label{fig.3}
\end{figure*}

\section{Correlating morphology and metallicity}
\label{sec:mzr_krot}
\eagle's $z=0$ MZR is plotted in Fig.~\ref{fig.1}, with galaxies binned by 
$\kappa_{\rm{co}}$, the fraction of stellar kinetic energy of the galaxy that is invested in ordered rotation. Below $M_\star\sim 10^{10}~{\rm M}_\odot$, dispersion-supported galaxies (low $\kappa_{\rm co}$, orange line) have higher O/H than rotationally supported galaxies (blue line) of the same $M_\star$. Also striking is that O/H increases with $M_\star$ for rotationally supported galaxies, but is almost independent of $M_\star$ for dispersion-supported galaxies. As a consequence, the trend between O/H and $\kappa_{\rm co}$ inverts above $M_\star\approx 10^{10}~{\rm M}_\odot$, with massive dispersion-supported galaxies having {\it lower} O/H than rotationally supported galaxies of the same mass. We find similar trends in the higher resolution simulation Recal-L025N0752 (inset of Fig.~\ref{fig.1}).

The results of Fig.~\ref{fig.1}, combined with the dependence of O/H on the SFR, sSFR and the gas fraction reported by DR17, suggest that $\kappa_{\rm co}$ may itself correlate with these other galaxy parameters. We examine this in Fig.~\ref{fig.2}. 
Independent of $M_\star$, the gas fraction (middle panel) and the specific star formation rate (right panel) both increase with $\kappa_{\rm co}$, being the increase of $f_{\rm g}$ the most pronounced for the lower $M_\star$ galaxies (blue line), whereas the increase of the sSFR is more evident in the {\it higher} $M_\star$ galaxies (purple line). 
The sSFR is a measure of the rate at which metals are produced, and $f_{\rm g}$ a measure of the size of the reservoir that dilutes those metals. Therefore, a consequence of these trends is that O/H {\it decreases} with increasing $\kappa_{\rm co}$ at low $M_\star$, whereas it {\it increases} for high mass galaxies (see left panel); at $M_\star\sim 10^{10}~{\rm M}_\odot$, O/H does not depend on $\kappa_{\rm co}$. These findings are consistent with the relation between colour and kinematics of \eagle\ galaxies studied by \cite{correa2017}.
Our results are also consistent with the observations by \citet{calvi2018} that
late-type galaxies have higher SFR and sSFR compared to S0 and elliptical galaxies of the same mass.

How these correlations arise is analysed in more detail\footnote{Similar results are obtained if using $D/T$ or $V/ \sigma$ instead of $\kappa_{\rm co}$.} in Fig.~\ref{fig.3}. At $M_\star<10^{10}~{\rm M}_\odot$, galaxies typically have low $\kappa_{\rm co}$, but there is a tail of galaxies with high $\kappa_{\rm co}$ and low O/H: this tail generates the anti-correlation between $\kappa_{\rm co}$ and O/H in Fig.~\ref{fig.2}. These outliers are also gas-rich and have unusually high stellar ellipticities, $\epsilon_\star$, and low triaxiality parameter $T$. At $M_\star>10^{10}~{\rm M}_\odot$, galaxies have usually a high value of $\kappa_{\rm co}$, but now there is a tail of galaxies with low $\kappa_{\rm co}$ and high $T$ that are typically more massive and have low O/H. At intermediate masses, $M_\star\sim 10^{10}~{\rm M}_\odot$, there is relatively little variation in $\kappa_{\rm co}$ or $\epsilon_\star$.

\begin{figure*}
\centering
\subfigure{\includegraphics[width=0.97\textwidth]{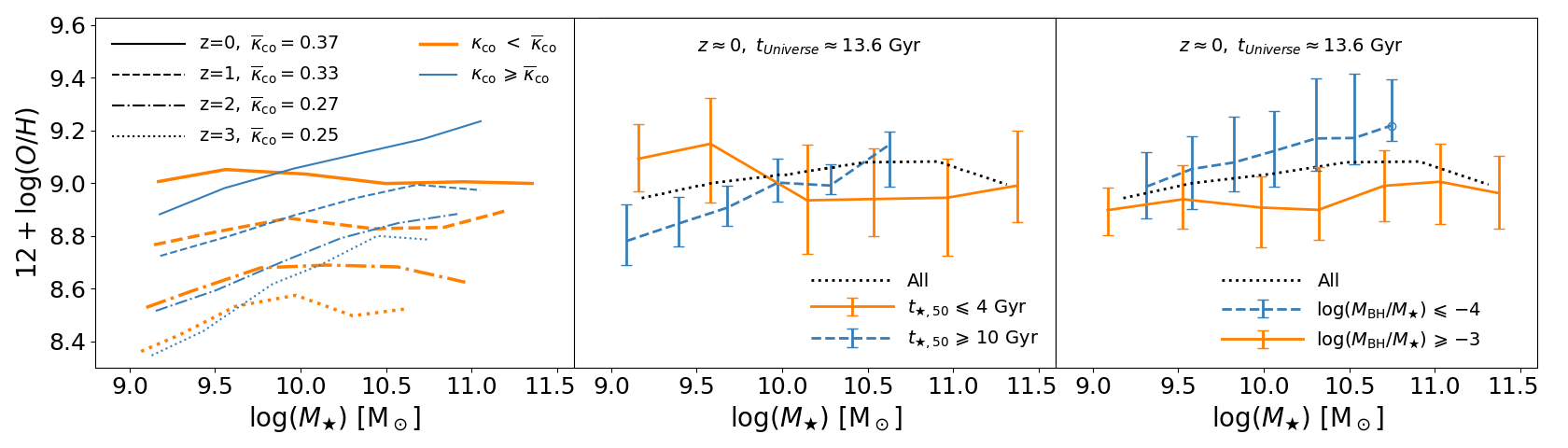}}
	\caption{ 
Left panel:
MZR at different $z$, as indicated in the figure. Curves representing galaxies with rotational support below the median
at the given $z$ are plotted in {\it thick orange}, those with higher $\kappa_{\rm co}$ are plotted in {\it thin blue}.
	Middle and right panels: $z=0$ MZRs binned according to ${t}_{{\star},50}$ (the time when galaxies reached half of their present $M_{\star}$) and $M_{\rm BH}/M_{\star}$ (the BH-stellar mass ratio), respectively. 
        }
\label{fig.4}
\end{figure*}

We plot the MZR relation at different redshifts $z$ in the left panel of Fig.~\ref{fig.4}. At a given $z$, the sample of simulated galaxies is separated in two sub-samples, using the median value of $\kappa_{\rm co}$ ($\bar{\kappa}_{\rm co}$) at that $z$. Note that the value of $\bar{\kappa}_{\rm co}$ tends to decrease with increasing $z$: dispersion-supported galaxies increasingly dominate at higher $z$. 
As expected, the normalization of the MZR decreases with $z$, with such evolution
being independent of $\kappa_{\rm co}$ at $M_{\star} \sim 10^{10}~{\rm M}_{\odot}$.
As we also noticed at $z=0$, there is a clear increase of O/H with $M_\star$ for galaxies
with high $\kappa_{\rm co}$, but this trend is mostly absent for the low $\kappa_{\rm co}$ galaxies.
At the low-mass end ($M_{\star} \la 10^{10}~{\rm M}_{\odot}$), 
the secondary dependence of O/H on ${\kappa}_{\rm co}$
tends to vanish as $z$ increases, due mainly to an increase of the MZR slope 
for systems with $\kappa_{\rm co} < \bar{\kappa}_{\rm co}$ at $z\ga1$.
On the other hand, at high masses ($M_{\star} \ga 10^{10}~{\rm M}_{\odot}$),
the secondary dependence of O/H on ${\kappa}_{\rm co}$ tends to be stronger 
at higher $z$, which, in this case, is caused by an increase of the MZR slope  
for systems with $\kappa_{\rm co} > \bar{\kappa}_{\rm co}$ at $z\ga 2$.
Similar evolutionary trends are obtained when using other morpho-kinematical indicators.

\section{Discussion}
\label{sec:discussion}

Reproducing observed metallicity scaling relations is an important test of current cosmological 
hydrodynamical simulations, with many of them showing good agreement with observations
(e.g., DR17, \citealt{dave2019, torrey2019}). 
In this context, metal-poor gas inflows have been suggested to play a key role on driving 
secondary O/H dependences (at a given mass) on galaxy sizes \citep[e.g.][]{Sanchez2018}, SFRs and 
gas fractions (e.g. DR17). 

In addition, \citet{torrey2018} claimed that
similar evolution timescales of SFR and O/H may be required to explain
the O/H-$M_{\star}$-SFR relation in the {\it{Illustris-TNG}} simulations.
For more massive galaxies, mergers and AGNs might also play a crucial role on shaping 
the MZR (e.g. DR17, \citealt{ma2016}). Although different works have focused on the O/H dependence on SFR, its relation with morpho-kinematics has not been discussed before. As a consistency check, we analysed the dependence of MZR on morpho-kinematics in the Illustris-TNG simulations, obtaining similar general trends to those shown in  Fig.~\ref{fig.1}: at low masses, galaxies with higher $D/T$ ratios generally have lower O/H, and the opposite is true for more massive systems. Hence, generally, such trends seem to be robust against the details of physical implementations adopted in these models. 
Nevertheless, we highlight that the detailed features of metallicity scaling
relations (e.g. detailed shape, scatter and normalization) do depend on the different model
prescriptions; this work is focused on predictions from the \eagle\ model.

\citet{Calura2009} analysed the MZR of galaxies using \lq chemical evolution\rq\ models.
They did not find a clear dependence of O/H on morphology at a given mass and $z$, which might
be a consequence of the different assumptions made in their models. For example, they assume that galaxies retain the same morphology throughout their evolution and they neglect mergers. In contrast in the \eagle\ simulations, discs may be destroyed during mergers and may re-grow following accretion \citep{Trayford2019}.

In the middle panel of Fig.~\ref{fig.4}, we show the \eagle\ $z=0$ MZR binned according 
to the cosmic time, ${t}_{{\star},50}$, when galaxies reached half of their present stellar mass, $M_{\star}$. At the low-mass end, systems with lower O/H tend to have been formed at later times. 
These systems also show higher sSFRs, higher gas fractions and lower stellar ages (DR17), exhibiting also 
higher rotational support and disky
morphologies (Fig.~\ref{fig.1} and ~\ref{fig.3}). An analysis of the formation histories
of these galaxies suggests that such trends would be associated with the accumulated effects of accretion
of metal-poor gas at {\it late} times (generally, $z\lesssim1$), 
which dilutes the metal content of galaxies, triggers their star formation activity and
contributes to the formation of the galaxy disc.\footnote{Long-term gas accretion could occur continuously or by successive gas inflow events.} 
This is consistent with the 
left panel of Fig. \ref{fig.4}, which shows that the dependence of the low-mass MZR on 
${\kappa}_{\rm co}$ is more significant towards $z=0$.

The right panel of Fig.~\ref{fig.4} shows the MZR binned according to the
black hole-stellar mass ratio ($M_{\rm BH}/M_{\star}$).  Massive galaxies with 
higher $M_{\rm BH}/M_{\star}$ tend to have lower O/H.  This is consistent with the results of DR17, who claimed that AGNs quench the metallicity evolution of galaxies by heating the gas,
suppressing the star formation activity and ejecting metals out of the systems.  Thus, at the high-mass end, less metal-enriched systems have lower sSFRs, lower fraction of star-forming gas and higher ages. Such systems tend also to be dispersion-supported, on average (Fig.~\ref{fig.1} and \ref{fig.3}). A preliminary analysis of the merger histories of \eagle\ galaxies suggests that massive galaxies with lower O/H seem to have been subjected to major mergers events, which inhibited the formation of a disc and also contributed to the increase of the central BH mass. The analysis of the formation histories of \eagle\ galaxies and its connection to the scatter of the MZR will be the subject of a forthcoming article (Zenocratti et al., in prep.).

\cite{sharma2019} propose a model of self-regulated galaxy formation, in which the star formation rate, stellar mass, gas mass and metallicity depend on halo mass, cosmological accretion rate and redshift, but regulated by feedback. In their \lq I${\kappa\epsilon\alpha}$\rq\ model, the main astrophysical parameter $\epsilon$ is a dimensionless measure of the efficiency of stellar feedback: a large value of $\epsilon$ implies efficient feedback and low star formation rate,  
while feedback is inefficient and the star formation rate is high for a small $\epsilon$. This leads to the following scaling for $Z$ and $f_{\rm g}$ with $M_\star$ and efficiency $\epsilon$:
\begin{equation}
Z \propto \frac{M_\star^{2/5}}{\epsilon^{3/5}}\,;\quad f_{\rm g}\propto  \frac{\epsilon^{\frac{2}{5}\frac{n-1}{n}}}{M_\star^{\frac{3}{5}\frac{n-1}{n}}}\approx \frac{\epsilon^{0.11}}{M_\star^{0.17}}\,,
\end{equation}
where $n\approx 1.4$ is the slope of the Kennicutt-Schmidt star formation law. These relations show that when feedback is efficient (meaning $\epsilon$ is large, $\epsilon\approx 1$), $Z$ is low and $f_{\rm g}$ is high, whereas on the other hand,
inefficient feedback ($\epsilon\ll 1$) implies $Z$ is high and $f_{\rm g}$ low.
We can connect this to the morphology of the galaxy by speculating that feedback in a thin disc - corresponding to a galaxy with high $\kappa_{\rm co}$ - is more efficient than when $\kappa_{\rm co}$ is low (a more spheroidal gas distribution). High-resolution simulations that resolve individual supernova energy injections in gas columns suggest this type of relation between feedback efficiency and disc morphology. For example, the simulations by \cite{Creasey2013} show that supernova explosions that occur above or below the disc are more efficient at driving winds, because cooling loses are suppressed when the explosion occurs at the lower densities that prevail there (see also \citealt{Girichidis2016}). Thus, in \eagle, accretion of metal-poor gas seems to trigger the formation of a disc but,
once the disc has already been formed, feedback effects would play a role on regulating 
the subsequent metallicity evolution. We leave the analysis of feedback effects on \eagle\ disky galaxies 
for a future work.

\section{Conclusions}
\label{conclusions}

We analysed the stellar mass-gas metallicity relation (MZR)  as function of morpho-kinematical parameters in the \eagle\ cosmological hydrodynamical simulations. At $z=0$, we found new secondary dependencies of metallicity on the internal kinematics and morphology of galaxies. At low masses ($M_{\star}\la 10^{10}~{\rm M}_{\odot}$), higher metallicities are found for more spheroidal and lower rotation-supported galaxies.

A preliminary analysis of the star formation histories of low-mass galaxies indicate that
late accretion of metal-poor gas dilutes the metal content
of galaxies, triggers their star formation activity and contributes to the formation of the galaxy disc.
When the gas is in a thin, rotationally supported disc, feedback may be more efficient, 
which results in lower O/H and a higher gas fraction \citep{sharma2019}. 
The trends in more massive galaxies are generally less strong, with lower metallicities found in more 
prolate galaxies with lower levels of rotational support. 
AGN feedback and mergers seem to play a key role on shaping the MZR at the high-mass end.
At increasing redshifts, the trend between rotational support and metallicity becomes weaker at low stellar mass but more pronounced at high stellar mass.
These trends are consistent with the secondary dependencies of O/H (at a fixed mass) on gas fraction, star formation rate and stellar age, and the relation between the latter quantities with galaxy morpho-kinematics (see DR17).

Our findings regarding the O/H secondary dependencies (at fixed stellar mass) on morpho-kinematics and
their relation with the scatter of the MZR were not previously discussed in the 
literature. 
A detailed analysis of the origin and evolution of the Mass-Metallicity-Morpho-kinematics Relation in \eagle\ will be presented in a future article.

\section*{Acknowledgements}
LJZ and MEDR acknowledge support from PICT-2015-3125 of ANPCyT, PIP 112-
201501-00447 of CONICET and UNLP G151 of UNLP (Argentina). MALL is a
DARK-Carlsberg Foundation Fellow (Semper Ardens project CF15-0384). We acknowledge the Virgo Consortium
for making their simulation data available. The EAGLE
simulations were performed using the DiRAC-2 facility at Durham,
managed by the ICC, and the PRACE facility Curie based in France
at TGCC, CEA, Bruy\`{e}res-le-Ch\^{a}tel.
This work used the DiRAC@Durham facility managed by the Institute for
Computational Cosmology on behalf of the STFC DiRAC HPC Facility
(www.dirac.ac.uk). The equipment was funded by BEIS capital funding
via STFC capital grants ST/P002293/1, ST/R002371/1 and ST/S002502/1,
Durham University and STFC operations grant ST/R000832/1. DiRAC is
part of the National e-Infrastructure.




\bibliographystyle{mnras}
\bibliography{biblio} 








\bsp	
\label{lastpage}
\end{document}